\documentclass{PoS}
\usepackage{subfigure}

\let\OLDthebibliography\thebibliography
\renewcommand\thebibliography[1]{
  \OLDthebibliography{#1}
  \setlength{\parskip}{0pt}
  \setlength{\itemsep}{0pt plus 0.0ex}
}

\def\beq{\begin{equation}}
\def\eeq{\end{equation}}
\def\bea{\begin{eqnarray}}
\def\eea{\end{eqnarray}}
\def\bq{\begin{quote}}
\def\eq{\end{quote}}
\def\ben{\begin{enumerate}}
\def\een{\end{enumerate}}
\def\bit{\begin{itemize}}
\def\eit{\end{itemize}}
\def\bfg{\begin{figure}}
\def\efg{\end{figure}}
\def\btp{\begin{tikzpicture}}
\def\etp{\end{tikzpicture}}
\def\bmp{\begin{minipage}}
\def\emp{\end{minipage}}

\makeatletter
\newcommand{\figcaption}[1]{\def\@captype{figure}\caption{#1}}
\newcommand{\tblcaption}[1]{\def\@captype{table}\caption{#1}}
\makeatother

\title{
{\vspace{-20mm}
{\normalsize\hfill{\small DESY 15-202}}}\\[10mm] 
Thermal evolution of the one-flavour Schwinger model using Matrix Product States}
\ShortTitle{Thermal evolution of the one-flavour Schwinger model using Matrix Product States}
\author{\speaker{H.~Saito}$^{a \star}$, M.~C.~Ba{\~n}uls$^b$, K.~Cichy$^{cd}$,  J.~I.~Cirac$^b$, K.~Jansen$^a$\\ 
    \llap{$^a$}John von Neumann Institute for Computing (NIC), DESY, Platanenallee 6, \\
    15738 Zeuthen, Germany\\
    \llap{$^b$}Max Planck Institute of Quantum Optics, Hans-Kopfermann-Str. 1, 85748 Garching, Germany\\
    \llap{$^c$}Goethe-Universit{\"a}t Frankfurt am Main, Institut f{\"u}r Theoretische Physik, \\
    Max-von-Laue-Str. 1, 60438 Frankfurt am Main, Germany\\
    \llap{$^d$}Faculty of Physics, Adam Mickiewicz University, Umultowska 85, 61-614 Pozna\'{n}, Poland\\
    E-mail: \email{saitouh@ccs.tsukuba.ac.jp}\\  \\
    $^\star$Current address: Center for Computational Sciences, University of Tsukuba, Tsukuba, Ibaraki 305-8577, Japan
}
\abstract{
The Schwinger model, or 1+1 dimensional QED, offers an interesting object of study, both at zero and non-zero temperature, because of its similarities to QCD. In this proceeding, we present the a full calculation of the temperature dependent chiral condensate of this model in the continuum limit using Matrix Product States (MPS). MPS methods, in general tensor networks, constitute a very promising technique for the non-perturbative study of Hamiltonian quantum systems. In the last few years, they have shown their suitability as ansatzes for ground states and low-lying excitations of lattice gauge theories. We show the feasibility of the approach also for finite temperature, both in the massless and in the massive case.
}
\FullConference{The 33rd International Symposium on Lattice Field Theory\\
                 14 -18 July  2015\\
                 Kobe International Conference Center, Kobe, Japan}
\begin{document}

\section{Introduction}
\vspace*{-0.2cm}
\label{sec:Intro}
Lattice gauge theory is a non-perturbative approach to investigate 
properties of gauge theories, e.g. Quantum Chromodynamics~(QCD), the theory of the strong interaction. 
Monte Carlo (MC) simulations have been used extensively and have become the standard technique.
However, the so-called sign problem is a well-known obstacle in MC simulations of, among others, QCD at finite chemical potential. 
Several approaches have been developed recently in order to overcome it, but so far no definite solution has been found.
Our method, based on the Hamiltonian formulation of lattice gauge theories~\cite{Banks:1975gq}, uses tensor network (TN) techniques \cite{verstraete08algo,schollwoeck11age,orus}, introduced in the context of quantum information and condensed matter physics.
The TN approach is an efficient way to approximate quantum many-body states and does not rely on MC simulations, thus it does not suffer from the sign problem.
As such, it has proven to be useful in investigating properties of several condensed matter systems that could not be tackled with MC methods due to the sign problem.

In the last years, we have studied the application of TN methods 
to lattice gauge theory, in particular the Schwinger model as a testbed~\cite{Cichy:2012rw,Banuls:2013jaa, Banuls:2013zva, Saito:2014bda, Banuls:2015sta}. 
In one dimension, a particularly efficient TN approach is the Matrix Product States (MPS) ansatz.
We have applied it to perform numerical studies of the Schwinger model at zero and non-zero temperature, showing the feasibility of this approach and its ability to yield precise and well-controlled results.
Other studies employing TN methods have been recently reported in Refs.~\cite{Buyens:2013yza, 
Rico:2013qya, Kuhn:2014rha, Kuhn:2015zqa, Shimizu:2014fsa, Shimizu:2014uva}.

In this proceedings, we show results of our calculations of the temperature dependence of the chiral condensate in the massless and massive Schwinger model.

\vspace*{-0.2cm}
\section{The Schwinger model for $N_{\rm f}=1$}
\vspace*{-0.2cm}
\label{sec:Schwgr}
We investigate the one-flavour Schwinger model, i.e. Quantum Electrodynamics in 1+1 dimensions.
One of the features that it shares with QCD is chiral symmetry breaking, although its mechanism, via the chiral anomaly, is different than in QCD.
The order parameter of chiral symmetry breaking is the chiral condensate $\Sigma=\left\langle {\bar \psi}\psi \right\rangle$. 
In the massless case, the temperature dependence of $\Sigma$ was computed analytically by Sachs and Wipf~\cite{Sachs:1991en}:  
\bea
   \left\langle {\bar \psi}\psi \right\rangle 
   &=& 
         \frac{m_{\gamma}}{2\pi} e^{\gamma} e^{2I(m_{\gamma}/T)} 
   = 
         \left\{ 
            \begin{array}{cl}  
                \frac{m_{\gamma}}{2\pi} e^{\gamma}  & \hspace{1cm} {\rm for} \hspace{1cm}  T\rightarrow 0  \\
                2T e^{-\pi T/m_{\gamma}}                   & \hspace{1cm} {\rm for} \hspace{1cm}  T\rightarrow \infty, 
            \end{array}  
         \right.
\eea
where $I(x) = \int_0^{\infty} \frac{dt}{1-e^{x\cosh(t)}}$, 
$\gamma=0.57721\cdots$ and $m_{\gamma}=g/\sqrt{\pi}$.
The formula shows that chiral symmetry is broken at $T=0$ and it gets fully restored ($\Sigma=0$) only at infinite temperature. Moreover, chiral symmetry restoration is smooth, without a phase transition.

The Hamiltonian of the one-flavour lattice Schwinger model (in the staggered discretization) was derived and discussed in~\cite{Banks:1975gq}: 
\bea
  H
  &=& x \displaystyle \sum_{n=0}^{N-2}
  \left[ \sigma_n^+ \sigma_{n+1}^- 
  + \sigma_n^- \sigma_{n+1}^+ \right]
  +\frac{\mu}{2} \sum_{n=0}^{N-1} 
  \Big[ 1+ (-1)^n \sigma_n^z \Big]
  + \sum_{n=0}^{N-2} \left[ L(n) \right] ^2\\
  &\equiv& H_{hop} + H_m + H_g,
  \label{eq:H}
\eea
where $x=1/g^2a^2$, $\mu= 2m/g^2a$, $m$ denotes the fermion mass, 
$a$ is the lattice spacing, $g$ is the coupling and $N$ the number of lattice sites.
The gauge field, $L(n)$, can be integrated out using the Gauss law: 
\beq
   L(n+1) - L(n) = \frac{1}{2} \left[ (-1)^{n+1} + \sigma_{n+1}^z \right].
   \label{eq:Gausslaw}
\eeq
This leaves only $L(n)$ at the boundary as an independent parameter and we take $L(0)=0$, i.e. no background electric field.

A convenient basis for the problem is thus
$\left| s_0 s_1 \cdots \right\rangle$ ~\cite{Banuls:2013jaa},
where $s_n=\{\downarrow,\uparrow\}$ is the spin state at site $n$ and all the gauge degrees of freedom have 
been integrated out.

\vspace*{-0.2cm}
\section{Tensor Network approach for thermal calculations}
\vspace*{-0.2cm}
\label{sec:therml_calc}
The thermal expectation value of $\Sigma$ at a given inverse temperature $\beta\equiv1/T$ is:
\bea
   \left\langle {\bar \psi}\psi \right\rangle 
   &=& 
         \frac{{\rm Tr} \left[ {\bar \psi}\psi \rho(\beta) \right]}
         {{\rm Tr} \left[ \rho (\beta)\right]},
         \label{eq:sigma}
\eea
where the (unnormalized) thermal density operator $\rho(\beta) \equiv e^{-\beta H}$. 
To investigate the temperature dependence of the chiral condensate, one needs to compute $\rho(\beta)$ in the relevant range of $\beta$.
The operator $\rho(\beta)$ can be approximated in the TN description by \cite{hastings06thermal}:
\bea
   \rho(\beta)
   &\approx& 
         \sum_{\{i_k,j_k\}}{\rm Tr} \left[ M[0]^{i_0j_0} \cdots M[N-1]^{i_{N-1}j_{N-1}} \right] |i_0\ldots i_{N-1}\rangle\langle j_0\ldots j_{N-1}|.
\eea
Each tensor $M[s]^{i_sj_s}_{n_sn_{s+1}}$ 
at each site $s=0,\,\cdots,\,N-1$ has four indices; two physical indices $i_s, j_s=0,\cdots,d-1$
~(here, $d=2$) 
and two virtual indices $n_s=0,\,\cdots,\,D-1$ introduced by the TN approximation. 
The maximal value of the virtual index is named the bond dimension, $D$.
This type of TN approximation is based on Matrix Product States (MPS). 
To be more specific, the considered operator is a Matrix Product Operator (MPO) \cite{verstraete04mpdo,zwolak04mpo,pirvu10mpo}.  
The elements of tensors $M[s]^{i_sj_s}_{n_sn_{s+1}}$ are computed in the following way.

Using $e^{-\beta H} = (e^{-\frac{\beta}{M} H})^{M}$, we divide the inverse temperature $\beta$ into $M$ steps of width $\delta=\beta/M$. 
The operator $e^{-\delta H}$ needs to be approximated in order to apply it to the MPO ansatz.
We use a second order Trotter expansion, based on the decomposition $H=H_e+H_o+H_z$, where 
 the hopping part is $H_{hop}=H_e+H_o$ and $H_z$ contains the mass term and the long range effective terms coming from the gauge part.
The exponentials of $H_e$ and $H_o$ can be written exactly as MPOs with a small bond dimension, but the exponential
of $H_z$, which includes long range terms, has to be approximated. This can be done using a first order Taylor expansion, so that
each step is finally written as a product of five operators,
\bea
   \exp \left(-\delta H \right) 
   &\approx& 
         \exp\left(-\frac{\delta}{2} H_e\right) 
         \left[1-\frac{\delta}{2} H_z\right]
         \exp\left(-\delta H_o\right) 
         \left[1-\frac{\delta}{2} H_z\right]
         \exp\left(-\frac{\delta}{2} H_e\right).
   \label{eq:expH}
\eea
In practice, the last factor of one step can be applied together with the first factor 
of the next one, so that only four operators per step are needed. 
The dominant error in this approximation comes from the Taylor expansion and is of $\mathcal{O}(\delta)$. 
It can, however, be controlled by extrapolating from a few $\delta$ values to $\delta=0$.

To obtain an MPO approximation for the thermal density operator at a given temperature, we start from the 
infinite temperature limit, $\beta=0$, where $\rho(0)$ is the identity operator, which corresponds to an 
MPO with bond dimension one.
On this state, we act with the succession of terms in Eq.~(\ref{eq:expH}), and approximate the result of each operation by an MPO
which is found by minimizing the Frobenius norm of the difference between the exact and approximated operators.
These steps produce an MPO approximation for the thermal density operator at an inverse temperature $\delta$.
By iterating the procedure, we can increase the inverse temperature in steps of $\delta$, until the maximum desired value, which
we choose to be close enough to the limit $T=0$. 
After any given step, we can compute the expectation value of the chiral condensate, given by Eq.~(\ref{eq:sigma}).

Importantly, one of advantages of the TN approximation is that it can be fully controlled and improved systematically by changing the bond dimension $D$.
The MPO description becomes exact for a sufficiently large $D$, namely exponential in the system size. 
However, for many physical systems of interest, including the Schwinger model, a relatively small value of $D$
is enough to attain the desired precision in the desired observables.
Fig.~\ref{fig:D} shows an example of convergence in $D$ for the chiral condensate.

\vspace*{-0.2cm}
\section{Results}
\vspace*{-0.2cm}
\label{sec:rslt}
One of the aims of this project is to confirm feasibility of TN in thermal calculations
of lattice gauge theory. 
For this purpose, we computed the chiral condensate of the one-flavour 
Schwinger model at non-zero temperature and compared it 
with the continuum analytical result of Ref.~\cite{Sachs:1991en}. 
To get the continuum result from the TN approach, one needs to perform the following extrapolations: $D\rightarrow\infty$, $\delta\rightarrow0$, the infinite volume extrapolation ($N\rightarrow\infty$) and finally the continuum limit extrapolation ($x\rightarrow\infty$).
The extrapolations are carried out in the above specified order.
Here, we only show an example of the $D$-dependence of the chiral condensate (Fig.~\ref{fig:D}). For a comprehensive description of our extrapolations, we refer to our recent paper \cite{Banuls:2015sta}.

\vspace*{-0.1cm}
\subsection{Full physical basis approach}

The basis used for the MPO approximation spans the exact physical space~\cite{Banuls:2013jaa}, 
so that the results obtained following the procedure described above do not involve any truncation of the gauge sector.
In Fig.~\ref{fig:cond_inCL}, we plot the temperature dependence obtained for the chiral condensate in the continuum limit.
Our results are consistent with the analytical result for $g\beta\geq0.5$, showing that the TN approach works well. 
The discrepancy in the high temperature region is caused by large cut-off effects. 
Fig.~\ref{fig:cont_extrapol_gbeta0.4} shows the continuum extrapolation 
at $g\beta=0.4$. 
We perform the extrapolation by using two fit functions: 
(a) $\Sigma=\Sigma_{\rm cont}+\frac{a_1}{\sqrt{x}} \log(x)+\frac{b_1}{\sqrt{x}}$~(solid blue), i.e. linear in the lattice spacing with a logarithmic correction (referred to as linear+log), 
(b) $\Sigma=\Sigma_{\rm cont}+\frac{a_2}{\sqrt{x}} \log(x)+\frac{b_2}{\sqrt{x}}+\frac{c_2}{x}$~(dashed red), i.e. additionally with a term quadratic in the lattice spacing (quadratic+log).
We clearly see that the results from these fits differ considerably and that the lattice spacings are not yet fine enough to approach the analytical value (almost exactly zero) with good precision.
Even though our method already allowed to reach rather fine lattice spacings, corresponding to $x=65$ (this can be compared to typical lattice spacings reached in MC simulations of the Schwinger model, having $x\leq10$), the precision that can be reached makes us susceptible even to terms cubic in the lattice spacing.
However, a fit with a term $(1/\sqrt{x})^3$ included is not meaningful without going to larger $x$.
At such large required values of $x>100$, we would need very large system sizes to control finite volume effects and such a computation was beyond the scope of this project.
We therefore tried a modified approach, described in the next subsection.

\bfg[t] 
   \bmp{.48\columnwidth}
   \centering
   \includegraphics[width=5.9cm]{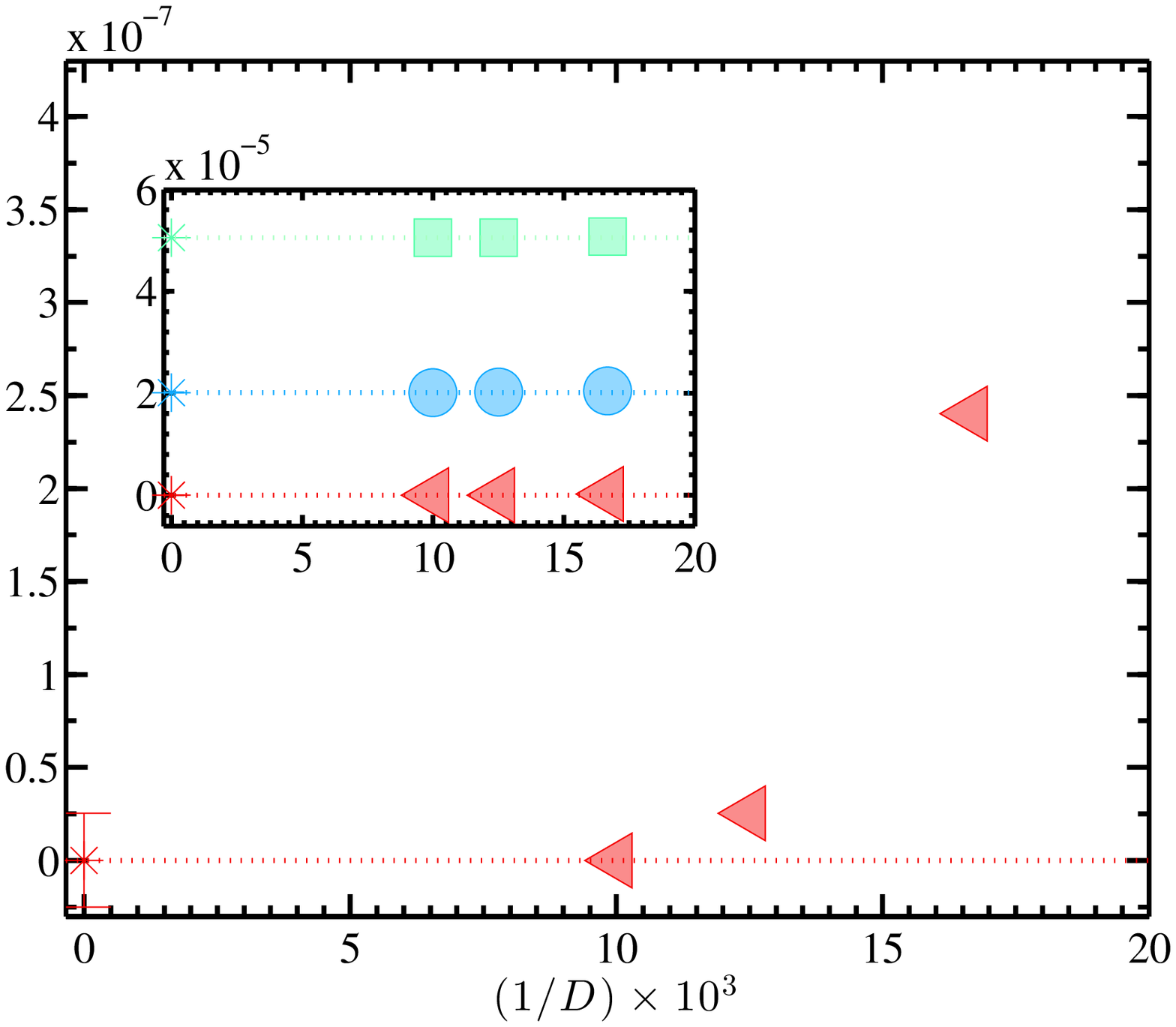}
   \caption{$D$-dependence of the chiral condensate for $\delta=5\cdot 10^{-4}$ (red), $7\cdot 10^{-4}$ (blue), $10^{-3}$ (green) 
   at $N=40$, $x=6.25$ and $g\beta=0.4$.}
   \label{fig:D}
   \emp
   \hspace{3mm}
   \bmp{.48\columnwidth}
   \centering
   \includegraphics[width=6.3cm]{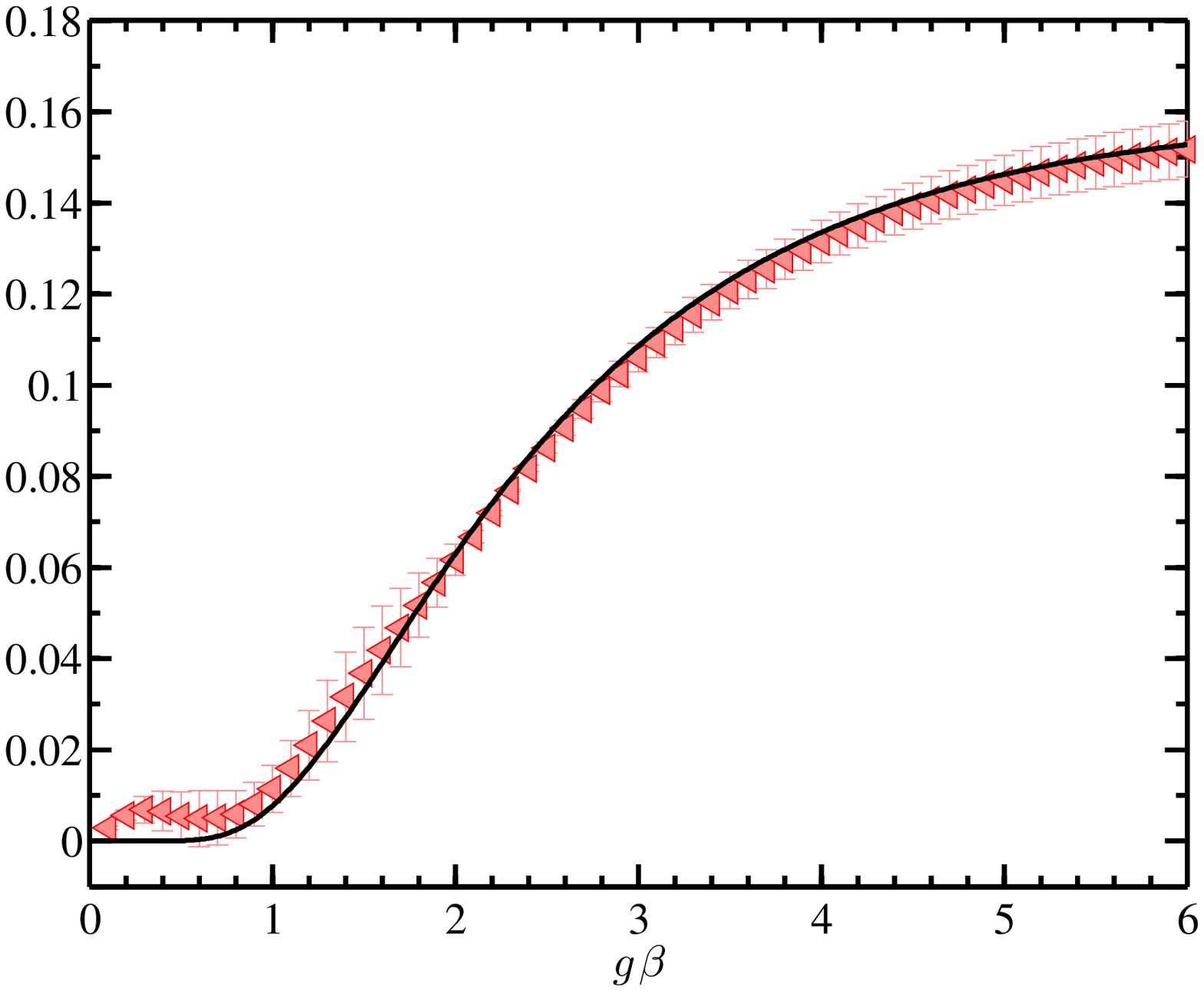}
   \caption{Chiral condensate in the continuum limit: 
                 computed by MPO approximation~(red symbols) 
                 and analytical result \cite{Sachs:1991en}~(black line).}
   \label{fig:cond_inCL}
   \emp
\efg

\vspace*{-0.15cm}
\subsection{Truncated gauge sector approach}
\label{sec:imp_app}

The Taylor approximation described in the previous paragraphs requires a small $\delta$ step for convergence. 
Since larger system sizes require the use of smaller $\delta$ steps, the computational cost severely increases for
larger system sizes, as the length of the chain has to increase as $N\propto \sqrt{x}$ to maintain a consistent physical volume.

An exact MPO expression  of the exponential $e^{-\delta H_z}$ is possible, with a bond dimension that scales as $N$
\cite{Banuls:2015sta}. 
Although such exact expression is not practical for large system sizes, it is possible to approximate it by an MPO truncated to a maximal bond dimension.
Such scheme, described in detail in the appendix of \cite{Banuls:2015sta},
corresponds to a truncation of the physical space to states where the absolute value of the electric field on each link does not exceed a maximum value, $l_{cut}$.

The gauge part of the Hamiltonian is diagonal in the spin basis. Then we can compute
$e^{-\delta \sum_{n=0}^{N-2} L(n)^2}\left| s_0 \cdots \right\rangle
=e^{-\delta \sum_{n=0}^{N-2} l_n^2}\left|  s_0 \cdots \right\rangle$,
where each value $l_n$ is determined from the spin (fermionic) content on all sites $i\leq n$ using Gauss' law (\ref{eq:Gausslaw}).
Thus, large values of $l_n$ are exponentially suppressed in the thermal state,
and hence a plausible assumption is to introduce a cut-off on the allowed values of $l_n$.
To be more precise, we project out states where $l_n \geq l_{\rm cut}$ for some $n=0,\cdots,N-1$.

This leads to high computational time savings if $l_{\rm cut}$ is taken to be small.
However, it can potentially lead to wrong results if $l_{\rm cut}$ is too small.
We checked several values of $l_{\rm cut}\leq16$ and concluded that a value of $l_{\rm cut}=10$ is enough for the range of parameters under consideration in this study.
With this approach, we could reach lattice spacings corresponding to an order of magnitude larger $x$ and hence perform the continuum limit extrapolations more reliably, at a similar computational cost as the full approach.
This not only makes the chiral condensate consistent with the analytical result for high temperatures, but also increases the precision for lower temperatures, see Fig.~\ref{fig:cond_inCL_athighT_Lcut}. 
Note that the approximation described in this subsection is consistent with another approach 
having an additional site for the gauge field~\cite{Buyens:2013yza}. 
\bfg[t] 
   \bmp{.48\columnwidth}
   \centering
   \includegraphics[width=6cm]{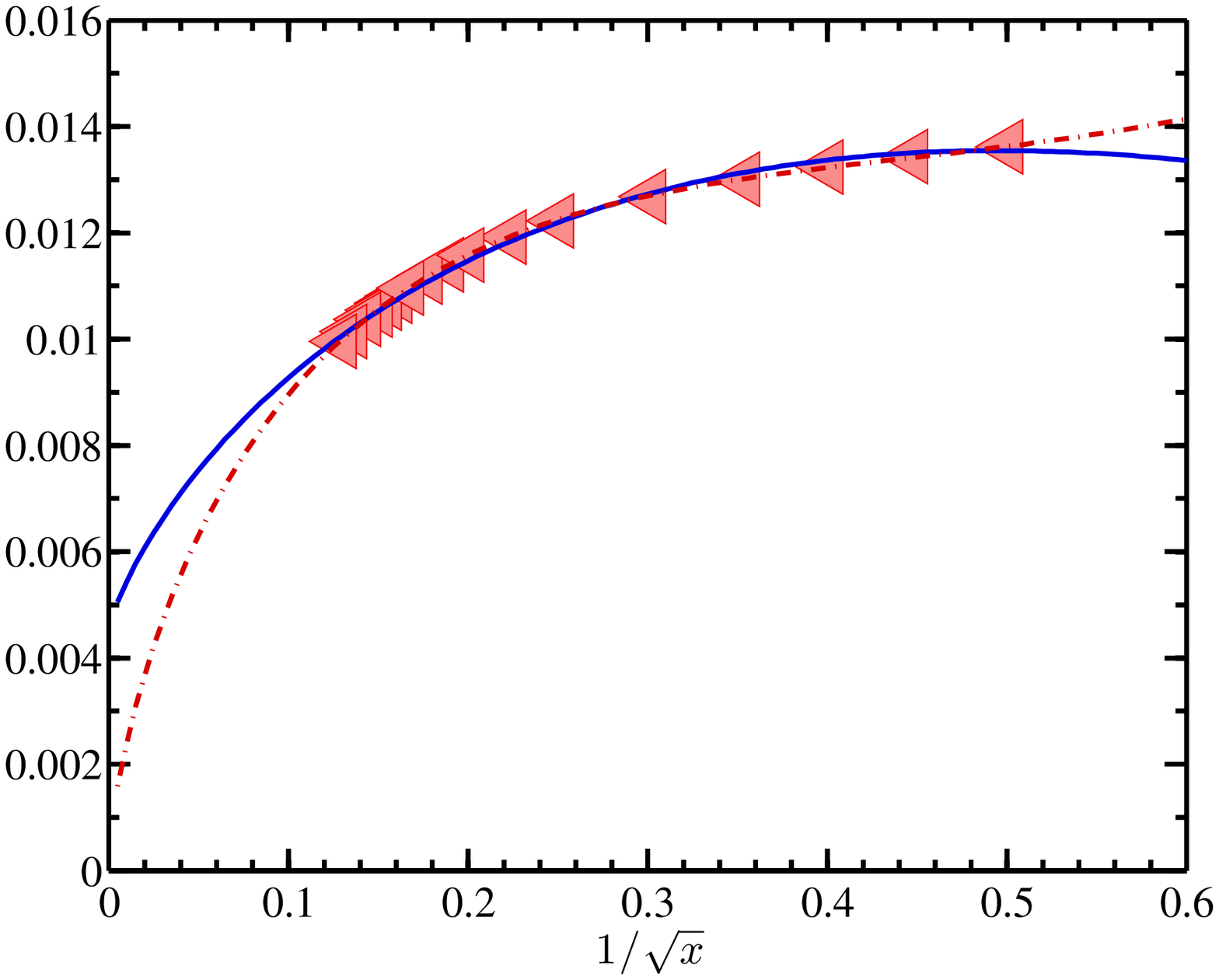} 
   \caption{Continuum extrapolation at $g\beta=0.4$, using two fit functions: linear+log and quadratic+log (for the form of the fit functions, see text).
    }
   \label{fig:cont_extrapol_gbeta0.4}
   \emp
   \hspace{3mm}
   \bmp{.48\columnwidth}
   \centering
   \includegraphics[width=6cm]{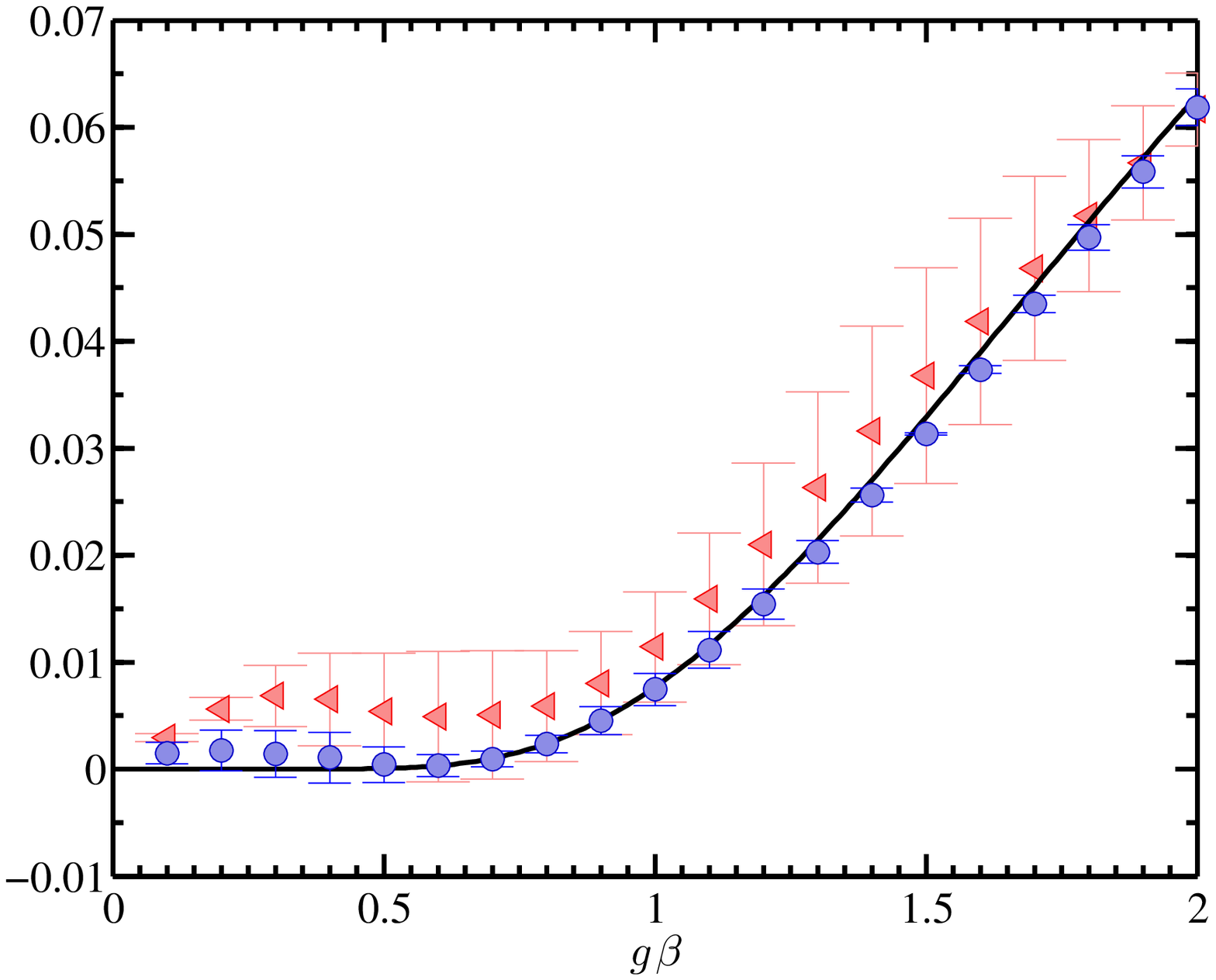} 
   \caption{Chiral condensate in the continuum limit with our full basis approach (red triangles) and with the truncated basis approach (blue circles).}
   \label{fig:cond_inCL_athighT_Lcut}
   \emp
\efg

\vspace*{-0.2cm}
\section{Further extension: Schwinger model in massive case}
\vspace*{-0.2cm}
\label{sec:massive}
Apart from investigating the exactly solvable massless case, our aim was to consider also the massive case, for which no exact solution is known.
Approximations valid for small fermion masses, however, exist, e.g. in Ref.~\cite{Hosotani:1998za}.
Our aim is to obtain the temperature dependence of the chiral condensate and thus show that the method works also in the massive case at $T>0$ and test the approximation of Ref.~\cite{Hosotani:1998za}.
Note that we already considered the $T=0$ massive case in Ref.~\cite{Banuls:2013zva}.

We shortly report here on our computation for the fermion mass $m/g=0.5$ using the truncated gauge sector approach -- see Fig.~\ref{fig:massive_demo}.
We show the $g\beta$-dependence of the subtracted chiral condensate. To obtain these values, we performed the following steps: (a) checks that $l_{\rm cut}=10$ is large enough, (b) extrapolation to infinite bond dimension, (c) extrapolation to $\delta=0$, (d) infinite volume extrapolation.
This led us to the bare (unsubtracted) infinite volume chiral condensate for several values of $x$.
Such an unsubtracted condensate contains a divergence that we subtracted using the zero-temperature chiral condensate in the free case.
The subtraction procedure is the same as described in Ref.~\cite{Banuls:2013zva} for the $T=0$ condensate.
Note that the divergence at the considered lattice spacings amounts to around 90\% of the interacting lattice value.

The main plot of Fig.~\ref{fig:massive_demo} shows how the $T=0$ continuum value, computed by us in Ref.~\cite{Banuls:2013zva}, is approached. Temperatures corresponding to $g\beta>4$ are effectively low enough such that the continuum extrapolation done at $g\beta=6$ should basically give the $T=0$ continuum value.
This is illustrated in the inset, where we show a fit of our lattice data (after all the extrapolations and divergence subtraction) using the linear+log fit function. Indeed, we obtain very good agreement with the expected continuum value (black square).
For more details about the thermal calculation for the massive Schwinger model, we refer to our forthcoming paper~\cite{OurLongerPaper}. 

\bfg[t]
   \centering
   \includegraphics[width=5cm, angle=270]{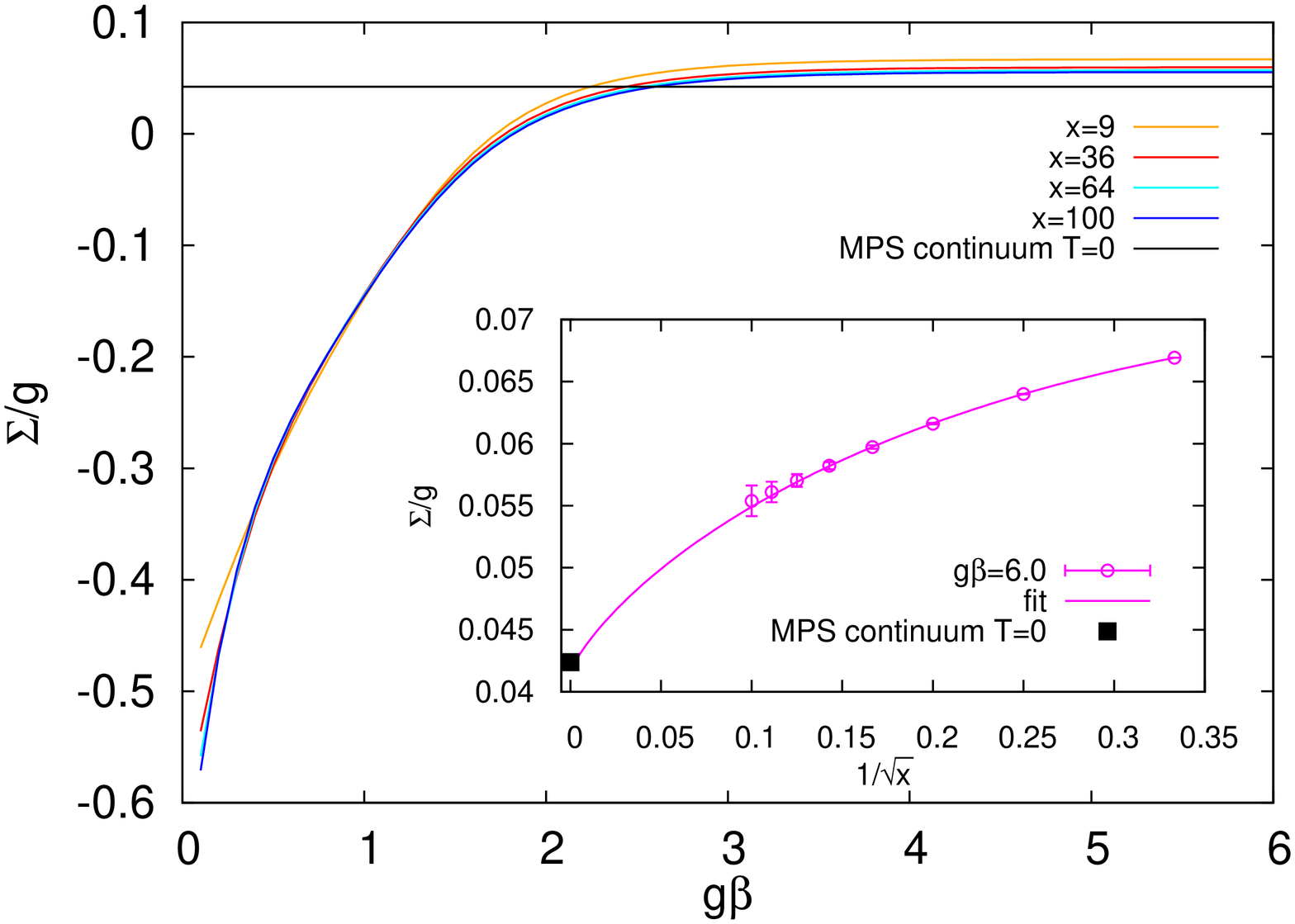} 
   \caption{Main plot: the temperature dependence of the chiral condensate for four selected lattice spacings. Shown is also the $T=0$ value from our calculation of Ref.~\cite{Banuls:2013zva}.
   Inset: continuum limit extrapolation using the linear+log fit ansatz. The black square shows the $T=0$ corresponding to the black line in the main plot.
   All the values of the chiral condensate are the subtracted ones after $D$, $\delta$ and $N$ extrapolations.}
   \label{fig:massive_demo}
\efg

\vspace*{-0.2cm}
\section{Conclusion and outlook}
\vspace*{-0.2cm}
\label{sec:cncl}
We computed the temperature dependence of the chiral condensate of the one-flavour Schwinger model using the Tensor Network approach.
Both for the massless and massive case, we obtain results compatible with expectations, thus demonstrating the feasibility of the method.
The TN approach is a prospective way of dealing with lattice gauge theories with a sign problem in MC simulations, in particular with QCD at non-zero chemical potential.
Although we have not yet addressed such theories, our research programme is an important step in this direction.
Over the past two years, we have shown that TN methods are able to yield precise and well-controlled results in lattice gauge theories at zero and non-zero temperature.
One of our directly next steps will be to address a case of a model where the sign problem is present in MC simulations.

\vspace*{1mm}
\noindent\textbf{Acknowledgments.} HS was supported by the Japan Society for the Promotion of Science for
Young Scientists. This work was partially funded by EU grant SIQS (FP7-ICT 2013-600645).
KC was supported in part by the Helmholtz International Center for FAIR within the
framework of the LOEWE program launched by the State of Hesse.
Calculations on the LOEWE-CSC high-performance computer of the Frankfurt University were conducted for this research. 


%
%
%

\vspace*{-0.25cm}
\begin{thebibliography}{99}
\vspace*{-0.2cm}
\bibitem{Banks:1975gq}
T.~Banks, L.~Susskind and J.~B. Kogut,
Phys.\ Rev.\ D {\bf 13}, 1043 (1976).


\bibitem{verstraete08algo}
F.~Verstraete, V.~Murg and J.~I.~Cirac,
Advances in Physics, {\bf 57}, 143 (2008).
    
\bibitem{schollwoeck11age}
U.~Schollw{\"o}ck,
Annals of Physics {\bf 326}, 96 (2011).

\bibitem{orus}
R.~Orus,
Annals of Physics {\bf 349}, 117 (2014).
    

\bibitem{Cichy:2012rw}
K.~Cichy, A.~Kujawa-Cichy and M.~Szyniszewski,
Comput.\ Phys.\ Commun. {\bf 184}, 1666 (2013).

\bibitem{Banuls:2013jaa}
M.~C. Ba\~{n}uls, K.~Cichy, K.~Jansen, and J.I. Cirac,
JHEP {\bf 1311}, 158 (2013).

\bibitem{Banuls:2013zva}
M.~C. Ba\~{n}uls, K.~Cichy, J.~I. Cirac, K.~Jansen and H.~Saito,
PoS(LATTICE 2013)332.

\bibitem{Saito:2014bda}
H.~Saito, M.~C. Ba\~{n}uls, K.~Cichy, J.~I.~Cirac and K.~Jansen,
PoS(LATTICE 2014)302.

\bibitem{Banuls:2015sta}
M.~C. Ba\~{n}uls, K.~Cichy, J.~I. Cirac, K.~Jansen and H.~Saito,
Phys.\ Rev.\ D {\bf 92}, 034519 (2015).

\bibitem{Buyens:2013yza}
B.~Buyens {\it et al.},
Phys.\ Rev.\ Lett. {\bf 113}, 091601 (2014).

\bibitem{Rico:2013qya}
E.~Rico, T.~Pichler, M.~Dalmonte, P.~Zoller and S.~Montangero,
Phys.\ Rev.\ Lett. {\bf 112}, 201601 (2014).

\bibitem{Kuhn:2014rha}
S.~K{\" u}hn, J.~I. Cirac and M.~C. Ba{\~ n}uls,
Phys.\ Rev.\ A {\bf 90}, 042305 (2014).

\bibitem{Kuhn:2015zqa}
S.~K{\" u}hn, J.~I. Cirac and M.~C. Ba{\~ n}uls,
JHEP {\bf 1507}, 130 (2015).

\bibitem{Shimizu:2014fsa}
Y.~Shimizu and Y.~Kuramashi,
Phys.\ Rev.\ D {\bf 90}, 074503 (2014).

\bibitem{Shimizu:2014uva}
Y.~Shimizu and Y.~Kuramashi,
Phys.\ Rev.\ D {\bf 90}, 014508 (2014).

\bibitem{Sachs:1991en}
I.~Sachs and A.~Wipf,
Helv.\ Phys.\ Acta {\bf65}, 652 (1992).

\bibitem{hastings06thermal}
M.~Hastings,
Phys.\ Rev.\ B {\bf 73}, 085115 (2006).

\bibitem{verstraete04mpdo}
F.~Verstraete, J.~J. Garc{\'{\i}}a-Ripoll, and J.~I. Cirac,
Phys.\ Rev.\ Lett. {\bf 93}, 207204 (2004).

\bibitem{zwolak04mpo}
M.~Zwolak and G.~Vidal, 
Phys.\ Rev.\ Lett. {\bf 93}, 207205 (2004).

\bibitem{pirvu10mpo}
B.~Pirvu, V.~Murg, J.~I.~Cirac and F.~Verstraete,
New Journal of Physics {\bf 12}, 025012 (2010) 

\bibitem{Hosotani:1998za}
Y.~Hosotani and R.~Rodriguez,
J.\ Phys.\ A{\bf 31}, 9925 (1998).

\bibitem{OurLongerPaper}
M.~C. Ba{\~ n}uls, K.~Cichy, J.~I. Cirac, K.~Jansen and H.~Saito,
\newblock {in preparation}.

\end{thebibliography}
\end{document}